\documentclass[aps,prl,twocolumn,showpacs,superscriptaddress,showkeys]{revtex4-1}

\usepackage{amssymb}
\usepackage{amsmath}
\usepackage{amsfonts}
\usepackage{graphicx}
\usepackage{color}
\usepackage{xspace}
\usepackage{ulem}
\usepackage{mathtools}
\usepackage{hhline}

\newcommand{\AddrOxford}{Rudolf Peierls Centre for Theoretical Physics, University of Oxford, Oxford OX1 3PU, United Kingdom}
\newcommand{\AddrCoimbra}{Univ Coimbra, Faculdade de Ci\^encias e Tecnologia da Universidade de Coimbra and CFisUC, Rua Larga, 3004-516 Coimbra, Portugal}

\begin{document}

\title{Evaporating primordial black holes, the string axiverse, and hot dark radiation}

\author{Marco Calz\`a}    \email{mc@student.uc.pt}\affiliation{\AddrCoimbra}
\author{John March-Russell} %\email{john.march-russell@physics.ox.ac.uk}
\affiliation{\AddrOxford}
\author{Jo\~{a}o G.~Rosa} \email{jgrosa@uc.pt}\affiliation{\AddrCoimbra}

\date{\today}

\begin{abstract}
We show that primordial black holes (PBHs) develop non-negligible spins through Hawking emission of the large number of axion-like particles  generically present in string theory compactifications. 
This is because scalars can be emitted in the monopole mode ($l=0$), where no angular momentum is removed from the BH, so a sufficiently large number of scalars can compensate for the spin-down produced by fermion, gauge boson, and graviton emission. 
The resulting characteristic spin distributions for $10^8$-$10^{12}$ kg PBHs could potentially be measured by future gamma-ray observatories, provided that the PBH abundance is not too small. 
This yields a unique probe of the total number of light scalars in the fundamental theory, independent of how weakly they interact with known matter.  The present local energy density of hot, MeV-TeV, axions produced by this Hawking emission can possibly exceed $\rho_{\rm CMB}$. Evaporation constraints on PBHs are also somewhat weakened.
\end{abstract}

%\pacs{} 

\maketitle

%%%%%%%%%%%%%%%%%%%%%%%%%%%%%%%%%%%%%%%%%%%%%%%%%%%%%%%%%%%%%%%%%%%%%%%%%%%%%%%%%%%%%%%%

Superstring theory is one of the leading candidates for a fundamental theory combining quantum gravity and the Standard Model (SM) of particle physics. The energy scale at which `stringy' effects become relevant is, however, unknown, and it may well be too high for these effects to be tested in the foreseeable future. The need for six additional compact spatial dimensions, with unknown size and geometry, and (broken) supersymmetry also leads to a plethora of different low energy theories, making it hard to identify specific signatures of the underlying fundamental theory that could be tested in the laboratory or with astrophysical observations.

There is, however, a fairly generic prediction of string compactifications if the strong CP problem is solved by the Peccei-Quinn mechanism \cite{Peccei:1977hh, Weinberg:1977ma, Wilczek:1977pj}, namely the existence of a large number of light pseudoscalar particles in the effective four-dimensional theory \cite{Arvanitaki:2009fg}.  These axion-like particles (ALPs) arise as the zero-modes in the Kaluza-Klein expansion of antisymmetric tensor fields in the underlying string theory, including the Neveu-Schwarz 2-form $B_2$, existent in all closed string theories, and the Ramond-Ramond $p$-forms of Type II and Type 1 string theories \cite{Svrcek:2006yi}. (There are also other ways that  ALPs can arise.) Each $p$-form can give rise to a multitude of ALPs in the 4d theory, as many as the number of homologically inequivalent $p$-cycles (closed surfaces) in the extra-dimensional manifold. This number is typically large, of the order of $10^2$ or even $10^5$, simply due to the number of different ways in which a closed surface can be embedded within a six-dimensional compact manifold. 

One of these particles can be the strong-CP-problem solving QCD axion, and much like the latter a large fraction of the ALPs are protected at all loop orders by shift symmetries inherited from the gauge symmetries of their parent tensor fields. The generic expectation is therefore that realistic string scenarios exhibit a large number of light or even ultra-light axions, whose exponentially small masses are generated solely by non-perturbative effects.

Light string axions can have a wide range of cosmological and astrophysical effects, e.g.~steps in the matter power-spectrum, rotation of the CMB polarization and black hole superradiant instabilities \cite{Arvanitaki:2009fg, Arvanitaki:2010sy}. These effects can yield interesting observational signatures of individual string axions in particular mass ranges, but not of the whole `string axiverse'. The number of relativistic degrees of freedom during cosmological nucleosynthesis or recombination could, in principle, be sensitive to the total number of light axions below a certain mass scale, but this depends crucially on whether they have been in thermal equilibrium with SM particles in the early Universe, which given the feebleness of their expected interactions with the latter is rather unlikely.

In this Letter, we propose a new way to probe the total number of  ALPs with mass $m < {\rm few}$~MeV through the spin distribution of primordial black holes (PBHs) that are evaporating today.  
PBHs \cite{Hawking:1971ei} can be formed in the early Universe through the gravitational collapse of large overdensities once their scale becomes smaller than the Hubble horizon \cite{Carr:1974nx, Carr:1975qj}. These could be generated by a plethora of different mechanisms, eg, non-standard inflation scenarios, curvaton models, or bubble collisions in cosmological phase transitions, spanning a wide range of masses. PBHs may account for at least a fraction of the dark matter in the Universe, which has sparked a renewed interest in this subject, alongside the recent detection of gravitational waves from a population of astrophysical BH binaries (see e.g.~\cite{Carr:2020xqk} for a recent review). 

Here we will be interested in small PBHs born with masses around $10^{11}$-$10^{12}$ kg, which are evaporating today. Although Hawking evaporation generically spins down a BH, an exception is the emission of light scalar particles, as first pointed out by Taylor, Chambers and Hiscock (TCH) \cite{Chambers:1997ai, Taylor:1998dk}, since these are the only type of particle that can be emitted in the monopole ($l=0$) mode. Scalar emission may therefore reduce a BH's mass without reducing its angular momentum, therefore increasing its dimensionless spin parameter $a_*=J/M^2$ (we set $G=\hbar=c=k_B=1$). In fact, TCH showed a BH evaporating solely through scalar emission would asymptote to a configuration with $a_*\simeq 0.555$, and that at least 32 light scalars could compensate the BH spin down through the emission of photons, neutrinos and gravitons.

Given the large number of ALPs expected in realistic string compactifications, this motivates exploring the spin evolution of PBHs in this context.  We will assume that Hawking emission is the only mechanism that affects the PBH spin, an assumption that we will  critically review at the end of our discussion. In the simplest scenario of PBH formation where large density fluctuations are generated on super-horizon scales and reenter the Hubble horizon during the radiation era, the nearly spherical collapse endows the BHs with only a small spin, at the few percent level \cite{Chiba:2017rvs, Mirbabayi:2019uph, DeLuca:2019buf}. There are, however, several possible scenarios where PBHs could be born with much larger spins, e.g.~formation during an early matter-dominated epoch \cite{Harada:2017fjm}. We will be agnostic about such initial conditions, since as we will show any natal spin is erased once a PBH has evaporated significantly {\it unless} a substantial number of light scalar fields is emitted.

To determine the evolution of PBHs we follow the formalism of Page \cite{Page:1976df, Page:1976ki}, where the dynamics of the BH mass and spin is determined by the functions $\mathcal{F}\equiv -M^2 dM/dt$ and $\mathcal{G}\equiv-(M/a_*) dJ/dt$, which removes the dependence on the BH mass. These are given by:
\begin{equation}\label{f_g}
\begin{pmatrix}
\mathcal{F}\\
\mathcal{G}
\end{pmatrix}=\sum_{i,l,m}\frac{1}{2\pi} \int_0^{\infty}dx \frac{\Gamma_{i,l,m}}{e^{2\pi k/\kappa}\pm 1}
\begin{pmatrix}
x\\
ma_*^{-1}
\end{pmatrix}
\end{equation}
where the sum is taken over all particle species $i$ and angular momentum quantum numbers $(l,m)$, $x=\omega M$, $k=\omega-m\Omega$ and $\kappa=\sqrt{1-a_*^2}/2r_+$ is the surface gravity of the Kerr BH, with $\Omega$ denoting its angular velocity at the event horizon, $r_+$. The upper (lower) sign in the denominator corresponds to fermion (boson) fields. The greybody factors $\Gamma=1-|Z_\mathrm{out}/Z_\mathrm{in}|^2$,
where $Z_\mathrm{in}$ and $Z_\mathrm{out}$ denote the amplitudes of the ingoing and outgoing wave modes, respectively. For massless fields, these can be computed by solving the Teukolsky equation for the radial wave function, upon separation of variables of the underlying wave equation(s), imposing ingoing boundary conditions at the horizon \cite{Teukolsky:1973, Press:1973zz, Teukolsky:1974yv}. This can also be done for massive fields, but given that the emission of particles with masses, $m$, above the Hawking temperature $T_H =\kappa/2\pi \simeq 1\, {\rm GeV} (10^{10}{\rm kg}/M)$ is exponentially suppressed we work in the approximation where particles are considered massless for $T_H>m$ and are otherwise absent from the emission spectrum.  As the methodology for computing these factors is well known, we refer the interested reader to e.g.~\cite{Taylor:1998dk} for more details.

We include in our numerical calculation an arbitrary number, $N_a$, of scalars which, for simplicity, we assume to have masses below the initial $T_H$, which is around the 10MeV scale for PBHs with a lifetime comparable to the age of the Universe. These are emitted alongside photons, gravitons, neutrinos and electrons/positrons from the start of our calculation. As a PBH evaporates, its $T_H$ increases, and we eventually include muons, taus and QCD degrees of freedom in the evolution at the corresponding mass thresholds. Pions ($\pi^0$ and $\pi^{\pm}$) are the only hadrons with masses well below the QCD scale, and so the only hadronic states included directly in the BH emission spectrum. For $T_H$ above the QCD scale a PBH emits elementary quarks and gluons that subsequently hadronize, and this can be taken into account by considering their effective QCD masses \cite{MacGibbon:1990zk}.  The effective masses given in \cite{ParticleDataGroup:2006fqo} are used; we have checked that our results do not change significantly if we use other values proposed in the literature, such as \cite{Iritani:2009mp}.

In Figure 1, we give our result for the present spin of PBHs, $a_{*0}$, as a function of their present mass, $M_0$, for different numbers of emitted ALPs, assuming an initial spin $a_*=0.01$ corresponding to PBH formation in the radiation era. Note that BHs in the given mass range correspond to the evaporated remnants of an initial population of PBHs with comparable masses and which are presently at different stages of their evolution. This justifies assuming a common initial spin. We also note that the critical PBH mass for a lifetime matching the age of the Universe of 13.8 Gyr exhibits a weak dependence on the number of different axions emitted, ranging from
$5\times10^{11}\,$kg in the absence of light axions to $\sim 2.7\times10^{12}\,$kg for $N_a=1000$, with a  $N_a^{1/3}$ scaling for $N_a\gtrsim 10$.

\begin{figure}[t]
\centering\includegraphics[width=0.85\columnwidth]{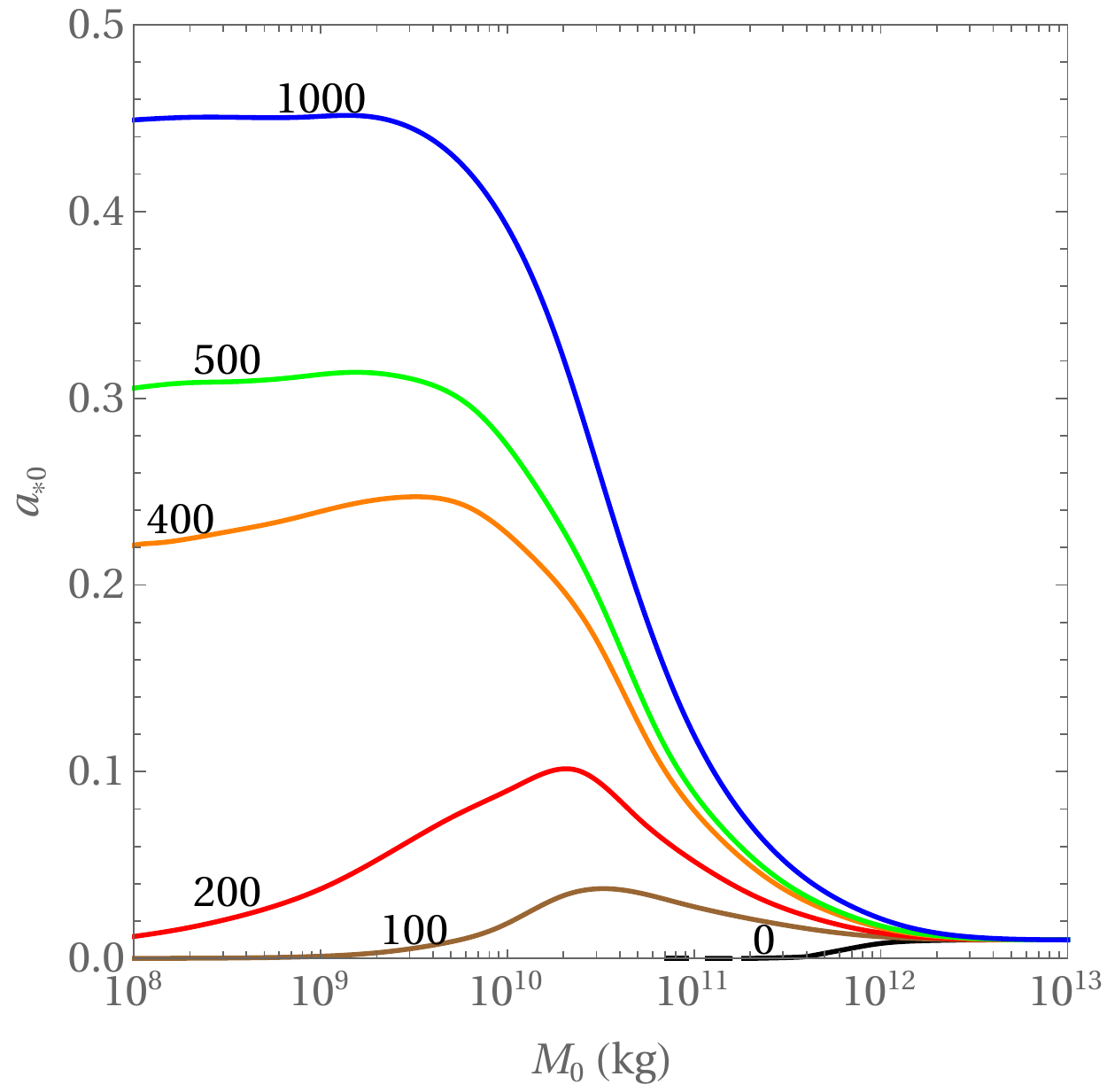}
\caption{Present PBH spin, $a_{*0}$, as a function of their present mass, $M_0$, for an initial population with spin $a_*=0.01$ and varying mass. Curves labelled by number of light ALPs.}
\end{figure}

While Figure 1 shows the current snapshot of the PBH Regge plot, it can also be viewed as a time evolution, since heavier BHs have evaporated little and lighter BHs have already lost a significant fraction of their original mass.  For $N_a=0$, PBHs quickly lose their initial spin as they evaporate, as has been the standard lore in the literature.  But the inclusion of many ALPs changes this picture considerably, with $a_*$ increasing initially due to the emission of the light scalars. For $N_a\lesssim 400$, the spin increases until $T_H$ reaches the threshold for quark and gluon emission, at which point the BH starts to spin down due to the large number of colored spin-1/2 and spin-1 degrees of freedom. For yet larger $N_a$, which is still plausible in string theories, the BH spin parameter stabilizes at values $a_{*0}\gtrsim 0.2$ and tending to the critical value $a_{*0}\simeq 0.555$ found by TCH for pure scalar emission. Note that, for $N_a\gtrsim 400$, PBHs never spin down completely even including all SM degrees of freedom, since $d\ln a_*/d\ln M=\mathcal{G}/\mathcal{F}-2$ has a non-trivial zero.

For comparison, we show in Figure 2 the results for an initial population of near-extremal PBHs.  Again, we find that even in this case PBHs with masses below $\sim 10^{11}\,$kg should have a presently negligible spin for $N_a=0$, but the inclusion of many light axions halts the BH spin-down or at least delays it considerably until QCD degrees of freedom can be efficiently emitted. Once more we find that for $N_a\gtrsim 400$, $a_{*0}\gtrsim 0.2$ for light BHs.
We note that the slightly oscillatory behaviour that can be observed in some of the curves in Figure 2 in the mass range $10^{10}$-$10^{11}$kg is due to the opposing effects of $\pi$'s and $\mu$'s on the spin evolution, with the former contributing to the BH spin-up and the latter spinning the BH down.

\begin{figure}[t]
\centering\includegraphics[width=0.85\columnwidth]{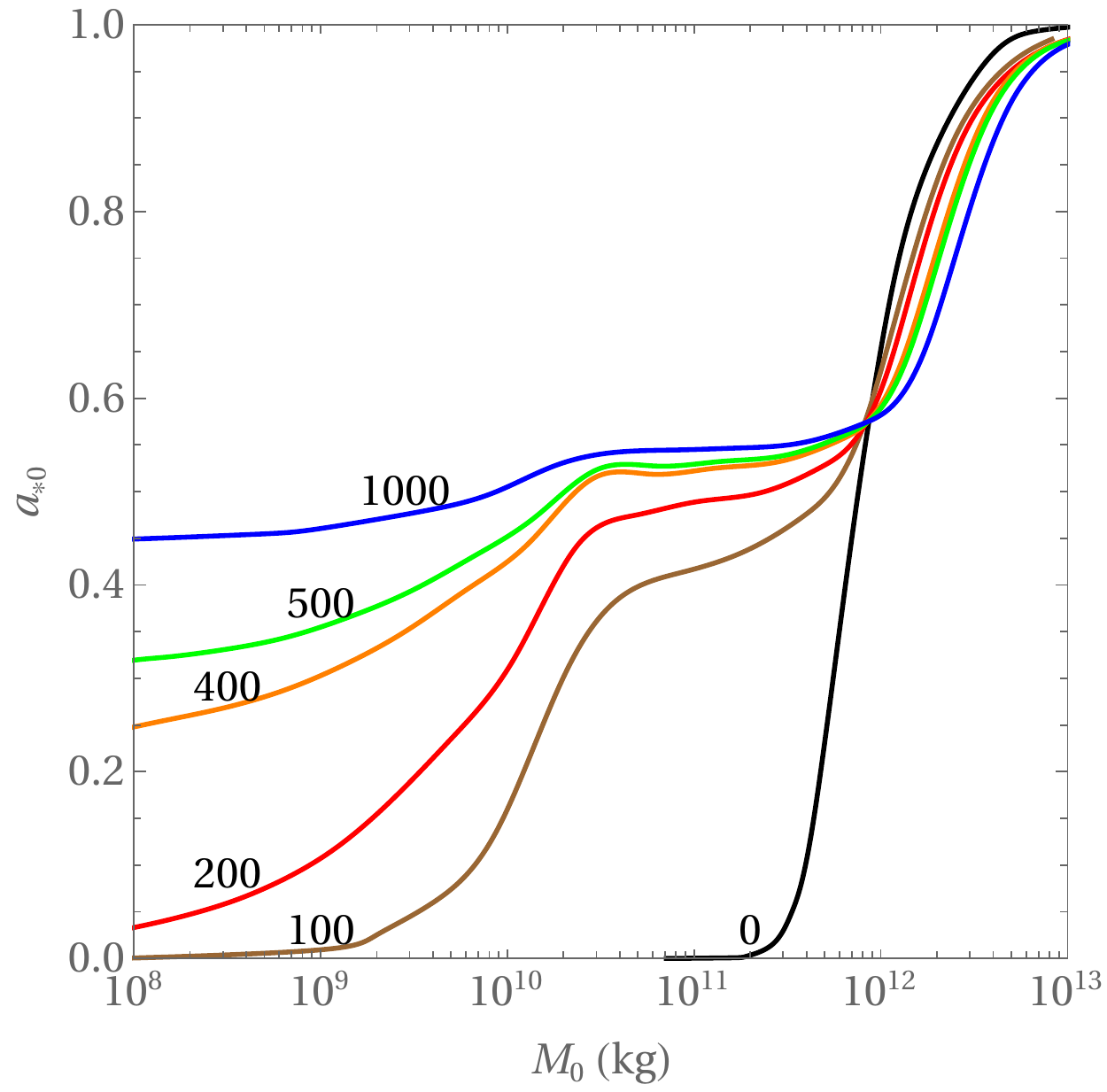}
\caption{As Figure 1 except initial spin parameter $a_*=0.99$.}
\end{figure}

Although not shown explicitly, we have found similar spin distributions for intermediate values of the initial spin, with either a `peak'- or `plateau'-like shape in the $10^{10}$-$10^{11}\,$kg (present) BH mass range, where $T_H\sim 0.1$-$1\,$GeV.  These are the PBHs that have evaporated essentially through the emission of non-colored degrees of freedom and for which the effect of the additional axions is more pronounced.  A generic feature is thus that PBHs in this mass range should have present spin parameters of at least $a_{*0}\sim {\rm few}\times 10^{-2}$, and up to or slightly over $\sim 0.5$ in the presence of hundreds of light axions, so that detecting BHs with such properties would constitute a `smoking-gun' for the string axiverse.

This conclusion is, of course, based on the assumption that evaporation is the main mechanism driving spin evolution. This is justified in the relevant PBH mass range as, firstly, accretion has been shown to be irrelevant for small PBHs with a lifetime close to the age of the Universe \cite{Rice:2017avg} and is also expected to occur in a quasi-spherical regime where a BH cannot efficiently spin up \cite{Ali-Haimoud:2016mbv}. 
Secondly, mergers should also be rare events for PBHs in this mass range. Following e.g.~\cite{Sasaki:2016jop}, two BHs can become gravitationally bound and decouple from the Hubble flow if their physical separation at matter-radiation equality $x<f^{1/3}\bar{x}$, where $\bar{x}$ is the average BH separation at this redshift and $f$ is the PBH fraction of the dark matter density. Given the existing constraints on $f$ for BHs that are evaporating today from the contribution of Hawking radiation to the extra-galactic diffuse gamma-ray background, $f\lesssim 10^{-7}$ \cite{Carr:2009jm, Arbey:2019vqx}, we conclude that only a small fraction of such BHs can become gravitationally bound and subsequently merge. Furthermore, mergers occurring early in the cosmic history will only affect the `initial' conditions for the subsequent evolution driven essentially by evaporation. 

Finally, PBHs in the interesting mass range also suffer from pion superradiant instabilities, where dense clouds of
$\pi^0$'s are produced around the PBHs at the expense of their rotational energy \cite{Ferraz:2020zgi}. However, such instabilities are only triggered for large initial spins and remain active for less than $\sim 1\,$Gyr, before the PBHs have evaporated significantly. Hawking evaporation will thus still
determine the final BH spins as assumed in our computation.
%Superradiant instabilities could also be triggered by other exotic particles, in particular if there are string axions with masses $\gtrsim$ MeV and sufficiently long lifetimes, potentially producing characteristic gaps in the primordial black hole mass-spin distribution [REF]. Since this is quite model-dependent and requires a simultaneous study of the evolution of the black holes' mass and spin through both superradiance and Hawking evaporation, we leave this analysis for future work.

The PBH spin distribution found here is, in any case, characteristic of the evaporation process in the presence of many light scalar fields and cannot, to our knowledge, be mimicked by other processes. Thus, observing such a distribution would constitute a unique signature of an underlying theory with a large number of light scalars.

Evaporating PBHs in the interesting mass range may in principle be detected through their Hawking photon emission, which includes both primary and secondary photons,  i.e.~those produced by the decays of the Hawking-emitted particles.  The latter dominate the flux for PBHs in the $10^{10}$-$10^{11}\,$kg mass range \cite{MacGibbon:2015mya}.  According to \cite{Fermi-LAT:2018pfs}, a $10^{10}\,$kg PBH could be detectable with Fermi-LAT up to a distance of $\sim 200\,$AU.  This compares to the mean distance between PBHs of present mass $M_0$ accounting for a present fraction $f_0$ of the local dark matter density $d \simeq 40\,{\rm AU} (M_0/10^{10}{\rm kg})^{1/3} (10^{-7}/f_0)^{1/3}$. 

Note the presence of many ALPs weakens the limits on the initial $f$ (and somewhat $f_0$) as, for a given present $M_0$ a reduced fraction of the total lifetime-integrated BH luminosity goes into photons.  Moreover the past-emitted photons are less energetic than they would have been in the absence of the ALPs as the past BH must be heavier, and thus colder, than in the $N_a=0$ case.  Precise evaluation of the new limits requires a dedicated study which we leave to future work.  (For related studies of modified PBH constraints in the context of extra-dimensional and other non-SM theories see \cite{Argyres:1998qn,Green:1999yh,Johnson:2020tiw,Baker:2021btk,Arbey:2021yke}.)  We take $f_0\sim 10^{-7}$
to roughly saturate current bounds \cite{Carr:2009jm, Arbey:2019vqx,footnote}.

Instead, the BH Hawking luminosity is dominated by a population of hot quasi-blackbody ALPs with mean energy $\sim 2.8 T_H$, and which are `dark' with respect to the SM up to possible feeble interactions set by $1/f_a$, where $f_a$ is the appropriate axion `decay constant'.  The integrated flux of ALPs from a single PBH is 
$\sim 3 \times 10^{22} N_a (10^{10}{\rm kg}/M)  s^{-1}$ in the relevant PBH spin range. 
Unlike the case of greatly-red-shifted dark matter particles and/or axions produced by the Hawking evaporation of micro PBHs in the very early pre-BBN universe \cite{Fujita:2014hha,Allahverdi:2017sks,Lennon:2017tqq,Hooper:2020evu} the axions in our case are hardly red-shifted at all as the relevant Hawking emission is occurring now.    

Although extremely challenging to detect, the much higher energies of these axions (and masses too for some of the ALPs) might enable new detection strategies as compared to the standard QCD axion solar flux expected for a given $1/f_a$.  Also note that the usual $\Delta N_{eff}$ constraints \cite{Lennon:2017tqq,Hooper:2019gtx,Hooper:2020evu} on extra relativistic degrees of freedom from BH evaporation derived from observations of BBN, CMB or structure-formation epochs do not apply
as the Hawking emission conversion of PBH mass into relativistic ALPs is dominated by late times 
and the present local energy density of hot `dark-radiation' ALPs can potentially exceed $\rho_{CMB}$.
Detection of one or more
sub-components of this background of energetic dark axions would be a striking signal for both axiverse physics
and the existence of Hawking evaporating PBHs. 

Returning to the photon flux, planned gamma-ray observatories, e.g.  HARPO and e-ASTROGAM, are expected to reach a somewhat better sensitivity than Fermi-LAT  in the energy range of 0.1-1GeV \cite{Knodlseder:2016pey}, although an improvement of a few orders of magnitude could be required if $f_0\ll 10^{-7}$, or if a precise measurement of the full photon spectrum, including secondaries, is required for an accurate determination of the PBH mass and spin.

An interesting possibility is that some ALPs could decay into photon pairs after emission.  Although this requires $f_a$'s well below the typical GUT scale values of string constructions, such reduced decay constants are possible
for some ALPs in warped and large-volume constructions \cite{Svrcek:2006yi,Flacke:2006ad,Cicoli:2012sz}.  In this case such ALPs would contribute to the secondary photon emission, which would encode additional information on the axiverse. The corresponding contribution to the diffuse gamma-ray background could also yield additional constraints on PBHs.

In Figure 3 we show the primary photon spectrum for different PBH masses and spins, given by
\begin{equation}
{d^2N_\gamma\over dt dE_\gamma}={1\over 2\pi}\sum_{l,m}{\Gamma_{1,l,m}(\omega)\over e^{2\pi \omega/\kappa}-1}~,
\end{equation}
with $E_\gamma=\omega$.  As one can see, the peak emission is quite sensitive to the PBH spin, while the corresponding energy depends mostly on the PBH mass.  A similar effect is obtained for spin-1/2 particles, so that we expect both the primary and secondary photon spectra to exhibit non-trivial dependences on the PBH mass and spin. Although we will leave a detailed analysis of the full spectrum for a dedicated future work, this suggests that in principle one can measure PBH spins $a_{*0}\gtrsim 0.1$ with at least a $\sim10\%$ precision measurement of the photon spectrum, provided the distance to the PBH can be measured with comparable accuracy through parallax.  Note that the time evolution of $T_H$, and thus photon flux and mean energy, is faster in the presence of many ALPs, than in the $N_a=0$ case. Given 
suitable high-precision measurements this can give confirmation of $N_a \neq 0$.

\begin{figure}[t]
\centering\includegraphics[width=0.8\columnwidth]{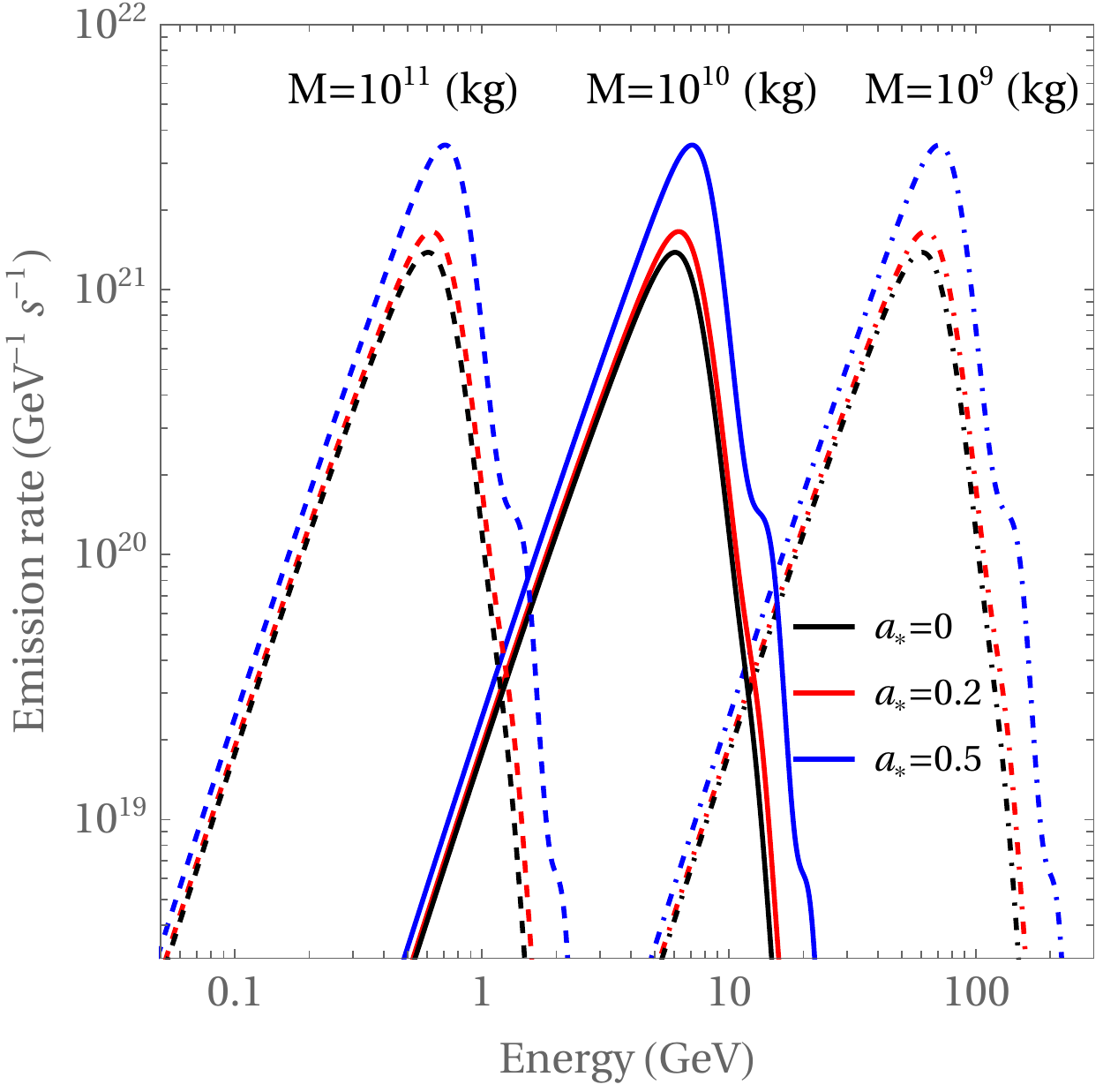}
\caption{PBH primary photon emission rate.}
\end{figure}

Moreover, halo PBHs velocities relative to the Earth are 2-3$\times 10^2\,$km/s, corresponding roughly to 40-60\hspace{0.025cm}AU$/{\rm yr}$.  Thus the population of bright (for both photons and hot ALPs) `point source'  evaporating PBHs in the solar neighbourhood can potentially change every few years, so increasing the chance of detecting BHs at different stages of their evaporation process.  This further allows one to fully assess the effects of a large number of axions.

In conclusion, detecting evaporating PBHs is of course a worthy endeavour to pursue on its own, given what it can tell us about both the early history of the Universe and the nature of semi-classical BHs. Our analysis shows that, in addition, their mass-spin distribution can give a unique probe of beyond the SM physics, which we hope may motivate future efforts in PBH detection.

\vspace{0.5cm} 
\begin{acknowledgments}

M.C. is supported by the FCT doctoral grant SFRH/BD/146700/2019. J.G.R.~is supported by the FCT Grant No.~IF/01597/2015 and by the CFisUC project No.~UID/FIS/04564/2019, and partially by the ENGAGE SKA grant (POCI-01-0145-FEDER-022217).

\end{acknowledgments}

\end{document}